\documentclass[fleqn]{article}

\usepackage{amsmath,amssymb}
\usepackage{graphicx}
\usepackage[top=0.75in, bottom=0.75in, left=0.75in, right=0.75in]{geometry}
\usepackage[small]{caption}
\usepackage[noblocks]{authblk}
\usepackage{hyperref}

 \def\tref#1#2#3{{#1} (#2) #3}
 \def\dsl#1{{#1}\!\!\!/}
\begin{document}

\title{{\sf Two-Loop QCD Renormalization and Anomalous Dimension of the Scalar Diquark Operator}}

\author[1]{R.T.\ Kleiv}
\author[1]{T.G.\ Steele}

\affil[1]{Department of Physics and
Engineering Physics, University of Saskatchewan, Saskatoon, SK,
S7N 5E2, Canada}

\maketitle

\begin{abstract}
The renormalization of the scalar diquark operator and its anomalous dimension is calculated at two-loop order in QCD, enabling higher-order QCD studies of diquarks. As an application of our result, the two-loop diquark anomalous dimension in the $\overline{{\rm MS}}$ scheme is  used to study the QCD renormalization scale dependence of diquark matrix elements of the $\Delta S=1$ effective weak Hamiltonian.    
\end{abstract}

\section{Introduction}
Four-quark (or tetraquark) $ qq\bar q\bar q$ states explain the inverted mass hierarchy of the
 scalar mesons  compared to a $q\bar q$ nonet in a variety of theoretical approaches \cite{jaffe,isgur,amir,maiani_scalars,forkel}.  With the inclusion of a gluonium (glueball) state \cite{glue}, the scalar spectrum below $2\,{\rm GeV}$ is then understood as mixtures of gluonium, the $q\bar q$ nonet, and the $qq\bar q \bar q$ nonet.  
The $X(3872)$ \cite{X3872} and $Y(4260)$ \cite{Y4260} mesons can also be interpreted as four-quark states \cite{maiani_xy}.  

Diquark ($qq$) clusters are relevant to the internal structure of hadrons (see e.g.,~\cite{diquark1,diquark2}). 
In particular, Ref.~\cite{maiani_xy}  uses constituent models for diquark clusters to study four-quark states.   The constituent (scalar) diquark masses that emerge in Ref.~\cite{maiani_xy} are in good agreement with  QCD sum-rule analyses of diquarks \cite{dosch_diquark,ailin}, providing QCD corroboration for the diquark model of four-quark states.

In this paper, we study the renormalization of scalar diquark operators to two-loop order in QCD and thereby obtain the two-loop anomalous dimension of the scalar diquark current.  As discussed below, the renormalization of the diquark operator is an essential component of QCD sum-rule analyses, and the anomalous dimension is also necessary for determining the scale dependence of matrix elements 
of the effective weak Hamiltonian for non-leptonic strange particle decays \cite{jamin}. Our two-loop results thus enable future QCD studies of diquarks to higher loop order.  

\begin{figure}[ht]
\centering
\includegraphics[scale=1]{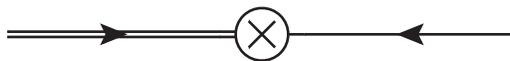}
\caption{Feynman diagram for the tree-level vertex of the diquark operator with the quark fields $\bar Q$ and $\bar q$. 
The double line represents the $Q$ field that is transposed and the diquark operator is  denoted by $\otimes$.
This and all subsequent Feynman diagrams were drawn with JaxoDraw \cite{jaxodraw}.
}
\label{feynman_rule}
\end{figure}

The scalar diquark operator in an anti-triplet colour configuration (the ``good" diquark in the terminology of Ref.~\cite{diquark2}) is given by \cite{dosch_diquark}
\begin{equation}
J_\gamma=\epsilon_{\alpha\beta\gamma}Q^\alpha_i \left(C\gamma_5\right)_{ij}q_j^\beta=
\epsilon_{\alpha\beta\gamma}Q^T_\alpha C\gamma_5q_\beta
 \,,
\label{diquark_current}
\end{equation}
where the greek and latin indices respectively represent colour and spin degrees of freedom for the quark fields $ Q$ and $q$, and $C$ is the charge conjugation operator. The presence of a transposed quark field in \eqref{diquark_current} implies that the Feynman rule for the three-point function of the diquark operator  and $\bar Q$, $\bar q$ fields shown in Fig.~\ref{feynman_rule} 
\begin{equation}
\Gamma^{(0)}_d=-\epsilon_{\alpha\beta\gamma}C\gamma_5,\,
\label{eq_feynman_rule}
\end{equation} 
implicitly transposes the external propagator associated with the $Q$ field. 

\section{One-Loop Renormalization}
\begin{figure}[ht]
\centering
\includegraphics[scale=0.5]{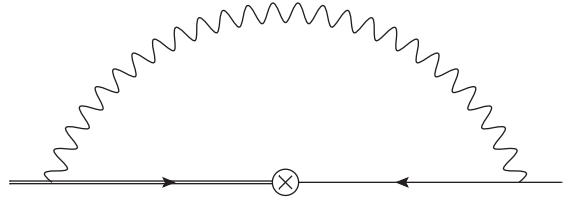}
\caption{One-loop Feynman diagram for the renormalization of $J_\gamma$. As in Fig.~\ref{feynman_rule},
the double line represents the (transposed) $Q$ field and the diquark operator is  denoted by $\otimes$.
}
\label{diquark_one_loop_diagram}
\end{figure}

\begin{figure}[ht]
\centering
\includegraphics[scale=0.6]{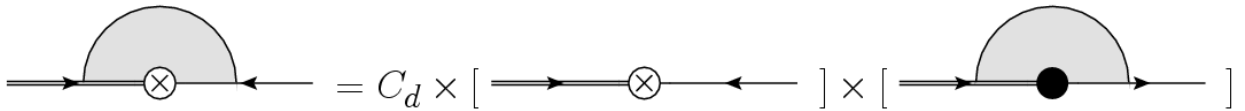}
\caption{Diagrammatic representation of the relationship \eqref{one_loop_relation} between two-point functions with scalar and diquark operator insertions. The  scalar operator is denoted by the solid circle.}
\label{relation}
\end{figure}

Although the diquark operator is gauge dependent, the theory  of composite-operator  renormalization \cite{composite} implies that 
the diquark operator is multiplicatively renormalizable because  
there are no lower-dimension operators with the same quantum numbers as \eqref{diquark_current}.\footnote{We are grateful for discussions with John Dixon clarifying this point.}  The one-loop renormalization of the diquark operator can thus determined by Fig.~\ref{diquark_one_loop_diagram}, which results in the following one-particle irreducible (1PI) Green function for a zero-momentum insertion of $J_\gamma$ in $D$-dimensions (dimensional regularization) 
\begin{equation}
\Gamma_d^{(1)}=i \frac{g^2}{4}\lambda^a_{\sigma\alpha}\lambda^a_{\tau\beta}\epsilon_{\sigma\tau\gamma}
\frac{1}{\nu^{2\epsilon}}\int \frac{d^Dk}{(2\pi)^D}\left(\gamma^\rho\right)^T \frac{\left(\dsl{p} +\dsl{k} \right)^T}{\left(p+k\right)^2}
C\gamma_5 \frac{\left(\dsl{p} +\dsl{k} \right)}{\left(p+k\right)^2}\gamma^\mu 
\left[-\frac{g_{\mu\rho}}{k^2} +(1-\xi) \frac{k_\mu k_\rho}{k^4}\right]\,,
\label{one_loop}
\end{equation}
where $\nu$ is the renormalization scale,  the quark mass has been ignored because dimensional regularization is a mass-independent scheme, 
$\alpha_s=g^2/(4\pi)$, colour indices have been explicitly shown for the Gell-Mann matrices $\lambda^a$, and a  covariant gauge with gauge parameter $\xi$ has been used.    
Working in normal (or naive) dimensional regularization,\footnote{We have chosen to work in normal dimensional regularization (as opposed to, e.g., the 't Hooft-Veltman scheme \cite{thooft}) because QCD sum-rule analyses of diquarks \cite{dosch_diquark,jamin} have used the  normal dimensional regularization scheme. } where $\left\{\gamma^\mu,\gamma_5\right\}=0$ \cite{chanowitz} in $D=4+2\epsilon$ dimensions,  and using the ($D$-dimensional) properties of the charge conjugation operator $CC=-1$ and $C\left(\gamma_\mu\right)^{T}C=\gamma_\mu$ \cite{delbourgo} we find
\begin{equation}
\Gamma_d^{(1)}=\frac{8}{3}\left[ -\epsilon_{\alpha\beta\gamma}C\gamma_5\right] i \frac{g^2}{4}
\frac{1}{\nu^{2\epsilon}}\int \frac{d^Dk}{(2\pi)^D}\gamma^\rho \frac{\left(\dsl{p} +\dsl{k} \right)}{\left(p+k\right)^2}
 \frac{\left(\dsl{p} +\dsl{k} \right)}{\left(p+k\right)^2}\gamma^\mu 
\left[-\frac{g_{\mu\rho}}{k^2} +(1-\xi) \frac{k_\mu k_\rho}{k^4}\right]\,.
\label{one_loop_2}
\end{equation}
By comparison with the one-loop process determining the renormalization of the scalar current $J_s=\bar Q q$, we see that \eqref{one_loop_2} can be related to the (one-loop) 1PI result for the scalar current $\Gamma^{(1)}_s$ apart from a numerical factor $C_d$ representing the ratio of the different colour factors that occur in the two processes
\begin{equation}
\Gamma_d^{(1)}=\frac{1}{2}\Gamma_d^{(0)}\Gamma_s^{(1)}\equiv C_d\Gamma_d^{(0)}\Gamma_s^{(1)}\,,
\label{one_loop_relation}
\end{equation}
as represented diagrammatically in Fig.~\ref{relation}.

The renormalized diquark operator $\left[J_\gamma\right]_R$ is defined via the renormalization constant $Z_d$, 
\begin{equation}
\left[J_\gamma\right]_R=Z_d J_\gamma\,.
\end{equation}
Similarly,  the well-known renormalization of the  scalar operator is 
\begin{equation}
\left[J_s\right]_R=Z_m J_s
\label{scalar_renorm}
\end{equation}
where $Z_m$ is the quark mass renormalization constant.  
Using \eqref{one_loop_relation} it is easy to see that to one-loop order in the minimal-subtraction (MS)  and associated schemes
\begin{equation}
Z_d=Z_{2F}^{1/2}Z_m^{1/2}\,,
\label{Zd_one_loop}
\end{equation}
where $Z_{2F}$ is the renormalization constant for the quark fields.
Landau gauge ($\xi=0$) is of particular interest in the QCD sum-rule analysis of diquark currents,   because  the Schwinger string used for a gauge-invariant formulation of the two-point diquark correlation function vanishes in this gauge \cite{dosch_diquark}.  Combining the one-loop Landau-gauge result $Z_{2F}=1$ with \eqref{Zd_one_loop} leads to the one-loop Landau gauge MS-scheme result
\begin{equation}
Z_d=Z_m^{1/2}=1+\frac{1}{2}\frac{\alpha}{\pi}\frac{1}{\epsilon}\,,
\label{one_loop_landau}
\end{equation}
where we use the dimensional regularization convention $D=4+2\epsilon$.  Eq.~\eqref{one_loop_landau} agrees with the (one-loop) renormalization and renormalization-group improvement implicitly implemented in Refs.~\cite{dosch_diquark,jamin}.

\section{Two-Loop Renormalization}
\begin{figure}[ht]
\centering
\includegraphics[scale=0.5]{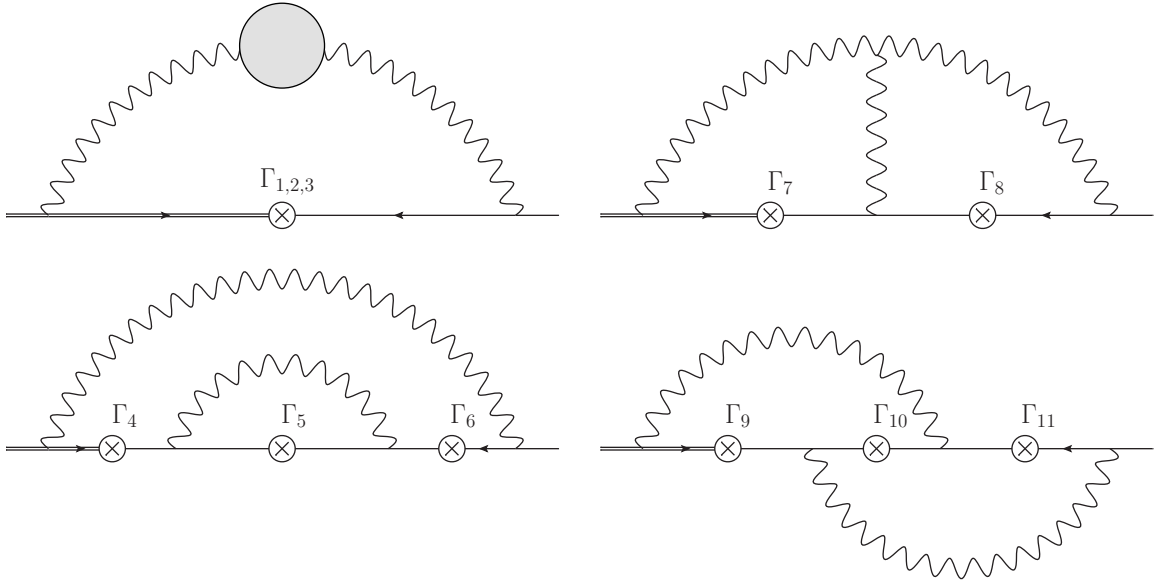}
\caption{Two-loop diagrams for the renormalization of the diquark operator where $\Gamma_1$ denotes a quark loop, $\Gamma_2$ a ghost loop and $\Gamma_3$ a gluon loop.  Implicitly, the $Q$ (double) line extends to the insertion of  the diquark operator.}
\label{diquark_two_loop_diagrams}
\end{figure}

 The two-loop diagrams for the renormalization of the diquark operator are shown in Fig.~\ref{diquark_two_loop_diagrams}.
As in the one-loop analysis and shown in Fig.~\ref{relation}, each diagram is given by a colour factor $C_d$ multiplying the bare diquark vertex and the equivalent diagram with a scalar current. The divergent parts  for each of the two-loop diagrams in Fig.~\ref{diquark_two_loop_diagrams} are expressed  in Table~\ref{results}  in terms of  the corresponding scalar diagram 
${\Gamma_{s,i}^{(2)}}$ 
in the modified minimal-subtraction ($\overline{\rm MS})$ scheme 
\begin{equation}
{\Gamma_{s,i}^{(2)}}=\left(\frac{\alpha_b}{\pi}\right)^2\left[\frac{A_i}{\epsilon}+\frac{B_i}{\epsilon^2}\right]\,, \quad i\in\left\{1\,,2\,,\,\ldots\,11\right\} \,,
\label{AB_definition}
\end{equation}
where   $n_f$ is the number of active quark flavours and $\alpha_b$ and $\xi_b$ are  the bare coupling and gauge parameter.  
A number of the Feynman diagrams are clearly related by the exchange of $Q$ and $q$ fields, and hence 
Table~\ref{results}  exhibits anticipated symmetries $\Gamma_4=\Gamma_6$, $\Gamma_7=\Gamma_8$ and $\Gamma_9=\Gamma_{11}$.
Note that the colour factors $C_d$ that relate the scalar and diquark diagrams are not universally equal to the one-loop result $C_d=1/2$, implying that one cannot expect the simple pattern of the one-loop result \eqref{one_loop_landau}  to persist at two-loop order.   The diagrams that are the exception to the one-loop pattern  ($\Gamma_5$ and $\Gamma_{10}$)  require  multiple applications of  colour algebra identities unique to the Feynman rule \eqref{eq_feynman_rule}; all other diagrams contain a single application of these identities combined with standard colour algebra factors occuring in the renormalization of the scalar operator.\footnote{In the previous version of this paper the Table~\ref{results} colour factor for diagram 10 in Fig.~\ref{diquark_two_loop_diagrams} was erroneous~\cite{john_gracey}.}

\begin{table}[ht]
\centering
  \begin{tabular}{|| l | l | l | l ||}
    	\hline
$i$	& $C_d$ & $A_i$ & $B_i$ \\ \hline & & & \\
	1 & $\frac{1}{2}$ & $\frac{n_f\left(2-L\right)}{6}$ & $-\frac{n_f}{12}$ \\ & & & \\
	2 & $\frac{1}{2}$ & $\frac{\left(2L-5\right)\left(1+\xi_b^2\right)}{32}$ & $\frac{1+\xi_b^2}{32}$ \\ & & & \\
	3 & $\frac{1}{2}$ & $\frac{\xi_b^2+4\xi_b-44-L\left(\xi_b^2+6\xi_b-25\right)}{16}$ & $\frac{25-6\xi_b-\xi_b^2}{32}$ \\ & & & \\
	4 & $\frac{1}{2}$ & $\frac{\xi_b\left[5+2\xi_b-L\left(3+\xi_b\right)\right]}{9}$ & $-\frac{\xi_b\left(3+\xi_b\right)}{18}$ \\ & & & \\
	5 & $\frac{1}{4}$ & $\frac{\left(3+\xi_b\right)\left[2L\left(3+\xi_b\right)-11-5\xi_b\right]}{18} $ & $\frac{\left(3+\xi_b\right)^2}{18}$ \\ & & & \\
	6 & $\frac{1}{2}$ & $\frac{\xi_b\left[5+2\xi_b-L\left(3+\xi_b\right)\right]}{9}$ & $-\frac{\xi_b\left(3+\xi_b\right)}{18}$ \\ & & & \\
	7 & $\frac{1}{2}$ & $\frac{3L\left(\xi_b^2+4\xi_b+3\right)-5\xi_b^2-17\xi_b-24}{16}$ & $\frac{3\left(\xi_b^2+4\xi_b+3\right)}{32}$ \\ & & & \\
	8 & $\frac{1}{2}$ & $\frac{3L\left(\xi_b^2+4\xi_b+3\right)-5\xi_b^2-17\xi_b-24}{16}$ & $\frac{3\left(\xi_b^2+4\xi_b+3\right)}{32}$ \\ & & & \\
	9 & $\frac{1}{2}$ & $-\frac{\left(3+\xi_b\right)\left[1+\xi_b\left(L-2\right)\right]}{72}$ & $-\frac{\xi_b\left(3+\xi_b\right)}{144}$ \\ & & & \\
	10 & $\frac{5}{2}$ & $\frac{3-6\xi_b-\xi_b^2}{144}$ & $0$ \\ & & & \\
	11 & $\frac{1}{2}$ & $-\frac{\left(3+\xi_b\right)\left[1+\xi_b\left(L-2\right)\right]}{72}$ & $-\frac{\xi_b\left(3+\xi_b\right)}{144}$ \\ & & & \\
    \hline
  \end{tabular}
\caption{Results for the two-loop diagrams in Fig.~\ref{diquark_two_loop_diagrams}. The quantity $L=\log(-p^2/\nu^2)$ and the notations for $A_i$ and $B_i$ are defined in Eq.~\eqref{AB_definition}.}
\label{results}
\end{table}

The two-loop renormalization procedure first involves the replacement of  $\alpha_b$ and $\xi_b$ with their (one-loop) renormalized expressions
(see, e.g., Ref.~\cite{pascual_tarrach_book}) 
\begin{gather}
Z_{\alpha}=1+\frac{\alpha}{\pi}\left[\frac{33-2n_f}{12\epsilon}\right] \,, \quad \alpha_b = Z_\alpha \alpha \,;
\label{Z_alpha}
\\
Z_{\xi}=1+\frac{\alpha}{\pi}\left[\frac{4n_f-39+9\xi}{24\epsilon}\right] \,, \quad \xi_b = Z_{\xi} \xi \,.
\label{Z_xi} 
\end{gather}
in the two-loop 1PI Green function 
\begin{equation}
\Gamma_d=\Gamma^{(0)}_d+\Gamma^{(1)}_d+\Gamma^{(2)}_d\,.
\label{gamma_full}
\end{equation} 
For consistency at two-loop level,  \eqref{gamma_full} requires inclusion of the finite parts of the one-loop calculation \eqref{one_loop_relation}
\begin{equation}
\Gamma^{(1)}_s=\frac{1}{3}\left(\frac{\alpha_b}{\pi}\right)\left[-\frac{3+\xi_b }{\epsilon }+2\left(2+\xi_b\right)-L\left(3+\xi_b\right)\right] \,, \quad L=\log{\left[-\frac{p^2}{\nu^2}\right]}\,.
\label{gamma_0_scalar} 
\end{equation}
The renormalization constant $Z_d$ is then constrained by the requirement that it cancel the divergences in 
\begin{equation}
Z_d Z_{2F}\left[ \Gamma^{(0)}_d+\Gamma^{(1)}_d+\Gamma^{(2)}_d \right]\,,
\end{equation}
where the two-loop $\overline{\rm MS}$ quark field renormalization constant is \cite{Z2F}
\begin{equation}
Z_{2F}=1+\frac{\alpha}{\pi}\frac{\xi}{3\epsilon}+\left(\frac{\alpha}{\pi}\right)^2 \left[\frac{\xi\left(27+17\xi\right)}{144\epsilon^2}+\frac{201-12n_f+72\xi+9\xi^2}{288\epsilon}\right] \,. 
\label{Z2F} 
\end{equation}
As a benchmark to ensure accuracy in our calculations in Table~\ref{results}, we have verified that our results for the scalar diagrams lead to the required two-loop $\overline{\rm MS}$ result $Z_s=Z_m$  \cite{Zm}
\begin{equation}
Z_{m}=1+\frac{\alpha}{\pi\epsilon}+\left(\frac{\alpha}{\pi}\right)^2 \left[ \frac{1}{\epsilon^2}\left( \frac{15}{8}-\frac{n_f}{12}\right)+\frac{1}{\epsilon}\left(\frac{101}{48}-\frac{5n_f}{72}\right)\right] 
 \,.
\label{Zm} 
\end{equation}
The final QCD result for the two-loop $\overline{\rm MS}$ 
diquark renormalization constant is 
\begin{equation}
Z_d = 1 + \frac{\alpha}{\pi}\left[\frac{3-\xi}{6\epsilon}\right] + \left(\frac{\alpha}{\pi}\right)^2 \left[\frac{1}{\epsilon}\left(\frac{1545-40n_f}{2880}-\frac{\xi}{8}-\frac{\xi^2}{64}\right) + \frac{1}{\epsilon^2}\left(\frac{234-12n_f}{288}-\frac{17\xi}{96}-\frac{5\xi^2}{288}\right) \right] 
 \,.
 \label{ZD}
\end{equation}
The cancellation of the $L/\epsilon$ terms in $Z_d$ that are generated by \eqref{gamma_0_scalar} provides another consistency check on our calculation. 
Note that the two-loop Landau gauge result does not uphold the one-loop ($\xi=0$) pattern $Z_d=Z_m^{1/2}$.

The anomalous dimension for the diquark operator defined by
\begin{equation}
\gamma_d=\frac{\nu}{Z_d}\frac{dZ_d}{d\nu}\,,
\label{anom_dim_def}
\end{equation} 
is easily extracted from \eqref{ZD} to obtain the two-loop $\overline{\rm MS}$ QCD anomalous dimension for the diquark operator
\begin{gather}
\gamma_d(\alpha)=\gamma_1\frac{\alpha}{\pi}+\gamma_2\left(\frac{\alpha}{\pi} \right)^2\, ,
\label{anom_dim}
\\
\gamma_1=1-\frac{\xi}{3}\,,\;\gamma_2=\frac{1545-40n_f}{720}-\frac{\xi}{2}-\frac{\xi^2}{16}\,.
\label{anom_dim2}
\end{gather}
In the extraction of the anomalous dimension we have verified that the two-loop coefficients of $Z_d$
\begin{equation}
Z_d=1+\frac{Z_{d,1}}{\epsilon}+\frac{Z_{d,2}}{\epsilon^2}+\ldots
\end{equation}
satisfy the renormalization-group constraint
\begin{equation}
2\alpha \frac{\partial Z_{d,2} }{\partial\alpha}=\left[
\gamma_d(\alpha)-\beta(\alpha)\alpha \frac{\partial }{\partial\alpha}-\delta(\alpha,\xi)\xi\frac{\partial}{\partial\xi}
\right] Z_{d,1}~,
\end{equation}
where we are working in the conventions of \cite{pascual_tarrach_book} with the (one-loop) $\beta$ function and anomalous dimension $\delta$ of the gauge parameter given by
\begin{gather}
\beta(\alpha)=\beta_1\frac{\alpha}{\pi}\, ,\;\beta_1=-\frac{11}{2}+\frac{n_f}{3}
\\
\delta(\alpha,\xi)=\delta_1\frac{\alpha}{\pi}\, ,\;\delta_1=\frac{1}{4}\left(13-3\xi\right)-\frac{n_f}{3}\,.
\end{gather}
Confirmation of this renormalization-group constraint provides another verification of the accuracy of our results given in Table~\ref{results}.

\section{Application and Conclusions}
It has previously been noted that at leading-order, the renormalization scale dependence cancels between the QCD perturbative contributions to the diquark decay constants and the $\Delta S=1$ effective weak Hamiltonian, although there remains some residual scale dependence from non-perturbative terms \cite{jamin}.  As an application of our two-loop results, we can explore this scale dependence at next-to-leading order.  Following Ref.~\cite{jamin}, we consider the combination 
\begin{equation} 
c_{-}(\mu) g_+(\mu)g_+(\mu)
\label{combo}
\end{equation}
where $c_{-}(\mu)$ represents the renormalization scale dependence of the Wilson coefficient in the $\Delta S=1$ effective weak Hamiltonian  \cite{buras} and $g_+(\mu)$ is the scale-dependent scalar diquark decay constant emerging from QCD sum-rules \cite{jamin}.   The renormalization-group (RG) factor arising from $c_-$ is \cite{buras}
\begin{equation}
c_-(\mu)\sim \exp{\left[-\int\frac{\gamma_-(\alpha)}{\beta(\alpha)}\frac{d\alpha}{\alpha}\right]}\,,
\end{equation}
where in the normal dimensional regularization scheme with  $n_f=3$, the anomalous dimension $\gamma_-(\alpha)$ is\footnote{Note that we have converted the expressions in \cite{buras} into our conventions.}
\begin{gather}
\gamma_-(\alpha)=\tilde\gamma_1\frac{\alpha}{\pi}+\tilde\gamma_2\left(\frac{\alpha}{\pi}\right)^2
\\
\tilde\gamma_1=-2\,,~\tilde\gamma_2=-\frac{50}{48}\,.
\end{gather}
Similarly, the anomalous dimension for the diquark operator leads to the following  RG factor for the (scalar) diquark decay constants
\begin{equation}
g_+(\mu)g_+(\mu)\sim \exp{\left[-2\int\frac{\gamma_d(\alpha)}{\beta(\alpha)}\frac{d\alpha}{\alpha}\right]}\,.
\end{equation}
As mentioned above, QCD sum-rule calculations with diquark currents extract gauge-invariant information from the two-point correlation function through the insertion of a Schwinger string, which becomes trivial for a line geometry in Landau gauge \cite{dosch_diquark}.  Thus for applications to RG behaviour of the diquark decay constants, we use \eqref{anom_dim2} with $n_f=3$ and $\xi=0$: 
\begin{equation}
\gamma_1=1,\,\gamma_2=\frac{95}{48}\,.
\label{two_loop_landau}
\end{equation}
The resulting RG behaviour of \eqref{combo} is
\begin{equation}
c_{-}(\mu) g_+(\mu)g_+(\mu)\sim
\exp\left[\int \frac{4}{9}\frac{\left[1+\frac{\tilde\gamma_2}{\tilde\gamma_1}\frac{\alpha}{\pi} \right]}{\left[1+\frac{\beta_2}{\beta_1}\frac{\alpha}{\pi}\right]} \frac{d\alpha}{\alpha}\right]
\exp\left[-\int \frac{4}{9}\frac{\left[1+\frac{\gamma_2}{\gamma_1}\frac{\alpha}{\pi} \right]}{\left[1+\frac{\beta_2}{\beta_1}\frac{\alpha}{\pi}\right]} \frac{d\alpha}{\alpha}\right]=1-\frac{35}{54}\frac{\alpha(\mu)}{\pi}~.
\label{run_result}
\end{equation}
Thus the leading-order cancellation of scale dependence in \eqref{combo} for the perturbative contributions to $g_+$ does not persist to second order.  However, the residual scale dependence associated with \eqref{run_result}, which decreases with increasing $\alpha(\mu)$, does have the right qualitative behaviour to counter the residual scale dependence encountered in Ref.~\cite{jamin}.  A more detailed analysis of the residual scale dependence is beyond the scope of this paper because it would require a full next-order sum-rule analysis of the diquark decay constants.

In conclusion, we have determined the $\overline{{\rm MS}}$ renormalization constant and associated anomalous dimension for the scalar diquark operator at two-loop order in QCD in an arbitrary covariant gauge for normal dimensional regularization.  
This result enables future QCD sum-rule studies of diquarks to higher-orders in perturbation theory.  
For example, the divergent terms in the diquark renormalization constant \eqref{ZD} combined with lower-loop ${\cal O}\left(\epsilon\right)$ and ${\cal O}\left(\epsilon^2\right)$ terms generate finite parts  corresponding to renormalization-induced physical contributions to the diquark  correlation function.  
Furthermore, 
 the anomalous dimension of the diquark operator appearing in the renormalization-group equation governing scale dependence of the diquark correlation function is an essential feature of QCD Laplace sum-rule analyses \cite{narison_rg}.
Given the relative size of the one- and two-loop terms in \eqref{ZD} and \eqref{two_loop_landau}, these renormalization-induced and anomalous dimension effects could be significant in higher-loop extensions of \cite{jamin}.

\bigskip
\noindent
{\bf Acknowledgements:}  We  are grateful for financial support
from the Natural Sciences and Engineering Research Council of Canada (NSERC).  We thank John Dixon and Derek Harnett for valuable discussions. We would also like to thank John Gracey for drawing our attention to an error in Table~\ref{results} in the previous version of this paper.


\end{document}